\documentclass[twocolumn]{aastex62}

\usepackage{color}

\shorttitle{LMC age gap clusters}
\shortauthors{Andr\'es E. Piatti}

\begin{document}

\title{Revisiting newly Large Magellanic Cloud age gap star clusters}

\author[0000-0002-8679-0589]{Andr\'es E. Piatti}
\affiliation{Instituto Interdisciplinario de Ciencias B\'asicas (ICB), CONICET-UNCUYO, Padre J. Contreras 1300, M5502JMA, Mendoza, Argentina}
\affiliation{Consejo Nacional de Investigaciones Cient\'{\i}ficas y T\'ecnicas (CONICET), Godoy Cruz 2290, C1425FQB,  Buenos Aires, Argentina}
\correspondingauthor{Andr\'es E. Piatti}
\email{e-mail: andres.piatti@unc.edu.ar}

\begin{abstract}

Recently, a noticeable number of new star clusters was identified in the
outskirts of the Large Magellanic Cloud (LMC) populating the so-called star cluster 
age gap, a space of time ($\sim$ 4-12 Gyr) where the only known star cluster is
up-to-date ESO\,121-SC\,03. We used Survey of the Magellanic Stellar History 
(SMASH) DR2
data sets, as well as those employed to identify these star cluster candidates, to produce relatively deep color-magnitude diagrams (CMDs) of 17 out 
of 20 discovered age gap star clusters with the aim of investigating them in detail. 
Our analysis relies on a thorough CMD cleaning procedure of the
field star contamination, which presents variations in its stellar density and
astrophysical properties, such as luminosity and effective temperature, around the star cluster fields. We built star cluster CMDs from stars with membership probabilities
assigned from the cleaning procedure. These CMDs and their respective spatial distribution maps favor the existence of LMC star field density fluctuations rather than age gap star clusters, although
a definitive assessment on them will be possible from further deeper photometry.
\end{abstract}


\section{Introduction} 

The absence of star clusters with ages between $\sim$ 3 and 10 Gyr in the
Large Magellanic Cloud (LMC) - the sole exception is ESO\,121-SC\,03
\citep{mateoetal1986}  - was noticed by \citet{oetal91}. They also found 
that the age gap correlates with a cluster metallicity gap, in the sense that star clusters 
younger than 3 Gyr are much more metal-rich that the ancient LMC globular 
clusters. The LMC age gap spans most of the galaxy lifetime, thus making difficult to
reconstruct its chemical enrichment from cluster ages and metallicities. Although different 
observational campaigns have searched for unknown old star clusters, they have confirmed
previous indications that star clusters were not formed during the age gap 
\citep[e.g.,][]{dc1991,getal97}.

The upper age limit of the LMC age gap is given by the youngest ages of the 15 LMC globular
clusters \citep[$\sim$ 12 Gyr,][]{pm2018,piattietal2018c}. The lower age limit, however, has
been changed as more intermediate-age stars clusters were studied in detail.
For instance, \citet{s98} found that NGC\,2121, 2155 and SL\,663 are $\sim$ 4 Gyr old
star clusters, while \citet{richetal2001} re-estimated their ages to be 0.8 Gyr younger
\citep[see, also,][]{petal02a}. Age estimates of poorly studied or unstudied star clusters
were derived during the last decade, and the oldest ones turned out to be $\sim$ 2.5-3.0
Gyr old \citep[see, e.g.][]{piattietal2009,p11a,pg13}. From the above results, we use here a 
conservative definition of age gap star clusters as those with ages between 4 and 12 
Gyr.

The LMC star cluster age distribution was modeled by \citet{bekkietal2004}, who
proposed that the LMC was formed at a distance from the Milky Way that did not 
allow its tidal forces to trigger star cluster formation efficiently. The star cluster
formation resumed in the LMC at its first encounter with the Small Magellanic 
Cloud $\sim$ 2-3 Gyr ago. Such a star cluster formation history
was not that of the Small Magellanic Cloud, which would have been formed
as a lower mass galaxy closer to the Milky Way, and thus more continuously
influenced by its gravitational field. Nevertheless, both Magellanic Clouds have
had a series of close interactions between them and  their first passage around
the Milky Way,
that  explain their abrupt observed chemical enrichment history and increase 
of the star cluster formation rates  \citep{p11a,p11b,pg13,kallivayaliletal13,lucchinietal2020}. 

Recently, \citet{gattoetal2020} performed a search for unidentified star clusters
in  LMC outermost regions and detected 20 star cluster candidates with estimated 
ages $\ga$ 4 Gyr. They used the YMCA (Yes, Magellanic Clouds Again) and STEP 
\citep[The SMC in Time: Evolution of a Prototype interacting late-type dwarf galaxy,][]{retal14} 
surveys carried out with the VLT Survey Telescope (VST) at ESO.
Because the discovery of only one LMC age gap star cluster
would be worth by itself, this astonishing large number of new age gap star cluster
candidates caught our attention. The authors mentioned that these candidates
come from surveying previously unexplored regions in the LMC periphery and
from their deep photometry. While the outermost LMC regions have started to be
targeted relatively recently with the aim of looking for new star clusters 
\citep[see Table\,1 in][]{maiaetal2019}, the depth of the photometric
campaigns could even improve \citep[see, e.g.,][]{s98,richetal2001}. In this work, we
make use of the same data sets employed by \citet{gattoetal2020} and
publicly available Survey of the Magellanic Stellar History
(SMASH) DR2 data sets \citep{nideveretal2021} to further confirm the ages of
the new star cluster candidates, so to reinforce their discoveries. In Section 2,
we present the data used in this work, while in Section 3 we describe the
analysis carried out in order to unveil the fiducial star cluster features in the
color-magnitude diagram (CMD). Finally, in Section 4 we discuss the present
results and summarize the main conclusions.

\section{The data}

\begin{figure}
\includegraphics[width=\columnwidth]{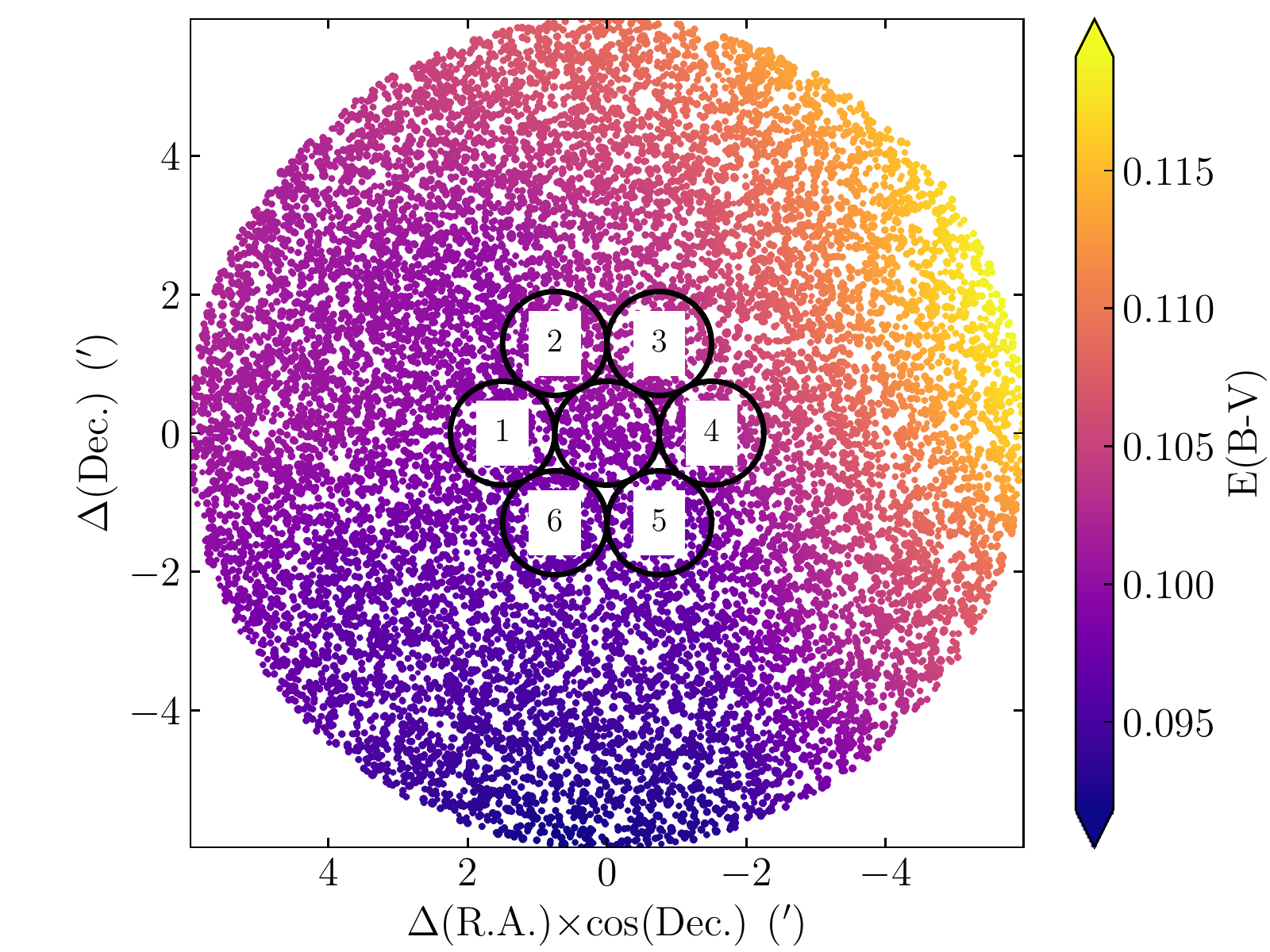}
\caption{Schematic chart centered on STEP-0029. The size of the symbols is
proportional to the $g$ brightness, while their {\bf color excesses are} coded according to the
color bar. The radius of the superimposed circles is 3 times that  adopted
as the cluster's radius \citep{gattoetal2020}. Six labeled reference star fields
distributed around the star cluster circle are also drawn.}
\label{fig1}
\end{figure}

We use the portal of the Astro Data Lab\footnote{https://datalab.noao.edu/smash/smash.php}, which is part of the Community Science and Data 
Center of NSF’s National Optical Infrared Astronomy Research Laboratory, to retrieve R.A and Dec. coordinates, PSF $g,i$ magnitudes and their respective errors, interstellar
reddening $E(B-V)$ and $\chi$ and {\sc sharpness} parameters of stellar sources
located inside a radius of 6$\arcmin$ from the star clusters' centers listed by \citet{gattoetal2020}. 
The retrieved data sets consist of sources with 0.2 $\le$ {\sc sharpness} $\le$ 1.0  and
$\chi^2$ $<$ 0.5, so bad pixels, cosmic rays, galaxies, and unrecognized double 
stars were excluded. \citet{gattoetal2020} discovered 85 star cluster candidates.
In this work we analyze 17 out of 20 objects with estimated ages $\ge$ 4 Gyr (log(age /yr)$
\ge$9.6). STEP-0004, YMCA-0031, and YMCA-0033 are age gap star cluster candidates, but 
they fall outside the areal coverage
of SMASH DR2. Figure~\ref{fig1} illustrates a typical star cluster field, where the variation 
of the interstellar reddening is shown with color-coded symbols.

The radii of the studied star cluster candidates  are relatively small, from 0.2$\arcmin$ up to 
0.55$\arcmin$, with an average of 0.35$\arcmin$  \citep[see Table B1 in][]{gattoetal2020}. Because we downloaded information
for circular areas much larger than the star cluster fields, we thoroughly monitored 
the  contamination of field stars in the star clusters' CMDs. Indeed, we selected for each
star cluster field, represented by a circle centered on the star cluster with a radius 3 times
that of the star cluster, 6 adjacent reference star field regions of equal star cluster field area
distributed around the
cluster region, as depicted in Fig.~\ref{fig1}. We based our analysis on dereddened
CMDs, so we first corrected by interstellar extinction the $g$ and $i$ magnitudes using the
$E(B-V)$ values provided by SMASH and the $A_\lambda$/$E(B-V)$ ratios, for $\lambda$ =
$g,i$, given by  \citet{abottetal2018}. The retrieved SMASH $E(B-V)$ value for each star
(see color bar in Fig.~\ref{fig1})
corresponds to the median $E(B-V)$ around it obtained by using the \citet{schlegeletal1998}'s 
reddening map \citep[see also][]{choietal2018,nideveretal2021}.

As for the data sets used by \citet{gattoetal2020}, they were kindly provided by V. Ripepi.
We note that  the $g,i$ bands of the STEP/YMCA surveys are not the same 
used by SMASH \citep[see figure 1 in][]{retal14}. Therefore, $g-i$ color ranges are not 
straightforwardly comparable, nor their CMDs. Both, STEP/YMCA and SMASH
data sets are then independent sources. \cite{nideveretal2017a} presented in their Table
4 the SMASH average photometric transformation equations. By adding in
quadrature zero-point, extinction and color term errors, we computed an 
accuracy $\la$ 0.02 mag in $gi$. They showed that these calibration errors imply a
 SMASH photometry precision of $\sim$ 0.5-0.7$\%$ in $gi$. Such a precision
 implies in turn an uncertainty of $\sim$ 0.11-0.14 mag in $gi$, for a star at
 the main sequence turnoff of a $\sim$ 4 Gyr old LMC star cluster
 ($g_0 \ga$ 22.0 mag). \citet{gattoetal2020} obtained an average photometry
 accuracy of 0.02-0.03 mag in $g$ and $i$, respectively, which is comparable to that
 of SMASH. We note that the zero point and color terms errors obtained by
 \citet[][see their Table 8]{retal14}, on which \citet{gattoetal2020}'s photometry relies,
are of the same order than those of SMASH.

\section{Data analysis}

The contamination of field stars plays an important role when dealing with star cluster
CMDs,  because it is not straightforward  to consider a star as a cluster member only
on the basis of   its position in that CMD. Sometimes, additional 
information like proper motions, radial velocities, and/or chemical abundances of individual 
stars can help with disentangling between field and star cluster members. Unfortunately, in 
the case of our  star cluster sample, {\it Gaia} DR2 proper motions  
\citep{gaiaetal2016,gaiaetal2018b} are unreliable at the main sequence turnoff level 
($g_0$ $\ga$ 22.0 mag). 
When such a piece of information is not available, photometry of reference star fields are
usually employed.
These reference star fields are thought to be placed far from the star cluster field, but not 
too  far from it as to become unsuitable as representative of the star field projected along the
 line-of-sight (LOS) of the star cluster. Frequently,  it is assumed that the stellar density and
the distribution of luminosities and effective temperatures of  stars in these 
reference star fields are similar to those of field stars located along the LOS of the star  cluster.
However, even though the
star cluster is not projected onto a crowded star field or is not affected by differential
reddening, it is highly possible to find differences  between the astrophysical properties of
the reference star fields and the star cluster field.
Bearing in mind the above considerations, we decided to clean the star field 
contamination in the star cluster CMDs by using, at a time, the 6 different devised reference 
star field areas introduced in Section 2.

The decontamination of a star cluster CMD comprises three main steps, namely: i)
to properly deal with each of the 6  reference star fields by considering 
the observed distribution of their stars in luminosity and effective temperature; ii) to
reliably subtract the reference star fields from the star cluster CMD 
 (one reference field at a time) and, iii) to assign
membership probabilities to stars that remained unsubtracted in the resulting cleaned star cluster CMDs. 
Stars with relatively high membership probabilities can likely be cluster members,
if they  are placed along the expected  star cluster CMD sequences. However, in general, they represent overdensities  along the LOS of the composite stellar  field population.
We refer the readers to \citet{pb12}, who devised the above procedure, which was satisfactorily
applied in cleaning CMDs of star clusters projected toward
crowded star fields  \citep[e.g.,][and references therein]{p17a,p17b,p17c} and affected by differential 
reddening \citep[e.g.,][and references therein]{p2018,petal2018}.

We  used a star cluster  field (a circle centered on the cluster with a radius 3 times that
of the cluster \citep[see Table B1 in][]{gattoetal2020} to subtract
a number of stars equal to that in  a reference star field.  
We repeated the star subtraction for the 6 devised reference star fields (see Fig.~\ref{fig1}), 
separately, one at a time. Note that if we subtracted less or more  stars than those
in the reference star field, we could
conclude on the existence of an unreal stellar excess or on a less populous aggregate,
respectively. Moreover, spurious overdensities  could even result if the cleaning procedure 
did choose the stars  to subtract following   a particular arbitrary spatial pattern 
(e.g., from North to South). If there were any intrinsic spatial gradient of field stars
in the cluster area, the  cleaning procedure would eliminate  it. The procedure finds 
field stars  in the star cluster CMD (with similar magnitudes and colors as those in the
reference star field CMD) where they actually are located, i.e.,
it subtracts more field stars where they are more numerous. 

The distribution of magnitudes 
and colors of the subtracted stars  from the star cluster field needs in addition to 
resemble that of the reference star field. The method consists in defining boxes centered
on the magnitude and color of each star of the reference star field CMD, then to superimpose
them on the star cluster CMD, and finally to choose one star per box to subtract.
With the aim of avoiding stochastic effects
caused by very few field stars distributed in less populated CMD regions, appropriate
ranges of magnitudes and colors around the CMD positions of field stars are 
advisable to be used. Thus, it is highly probable to find a star in the star cluster CMD
with a magnitude and a color within those box boundaries. In the case that more than one star is located inside that delimited CMD
region, the closest one to the center of that (magnitude, color) box is subtracted. In
the present work, we used boxes of ($\Delta$$g_0$, $\Delta$$(g-i)_0$) =
(2.0 mag,1.0 mag) centered on the ($g_0$, $(g-i)_0$) values of each reference field star.

In practice, for each reference field star, we first randomly  selected the position of a 
subregion inside the star cluster
field where to subtract a star. These subregions were devised as annular segments of 
 90$\degr$ wide
and of constant area. Their external radii are chosen
randomly, while the internal ones are calculated so that the areas of the annular
sectors are constant. Here we adopted an area  for the subregions equal 
to $\pi$$r_{cls}$, where $r_{cls}$ is the star cluster radius.  We then
looked for a star with ($g_0$, $(g-i)_0$) values
within a box defined as described above. If no star  is found in that annular sector,
we randomly  selected another one and  repeated the search,  allowing the
procedure to iterate up to 1000 times. If no star in the star cluster  field with a magnitude 
and a color similar 
to ($g_0$, $(g-i)_0$) is found  after 1000  iterations, we do not subtract any star 
 for that  ($g_0$, $(g-i)_0$) values. The same procedure was applied for all the stars
in the reference star field.
The photometric errors  of the stars in the star cluster field were also taken into account 
while searching for a star to be subtracted from the star cluster CMD. With that purpose, we 
iterated up to 1000 times  the search within each defined box, allowing the stars in the
star cluster CMD to vary their magnitudes and colors within an interval of $\pm$1$\sigma$,
where $\sigma$ represents the errors in their magnitude and color, respectively.

\begin{figure}
\includegraphics[width=\columnwidth]{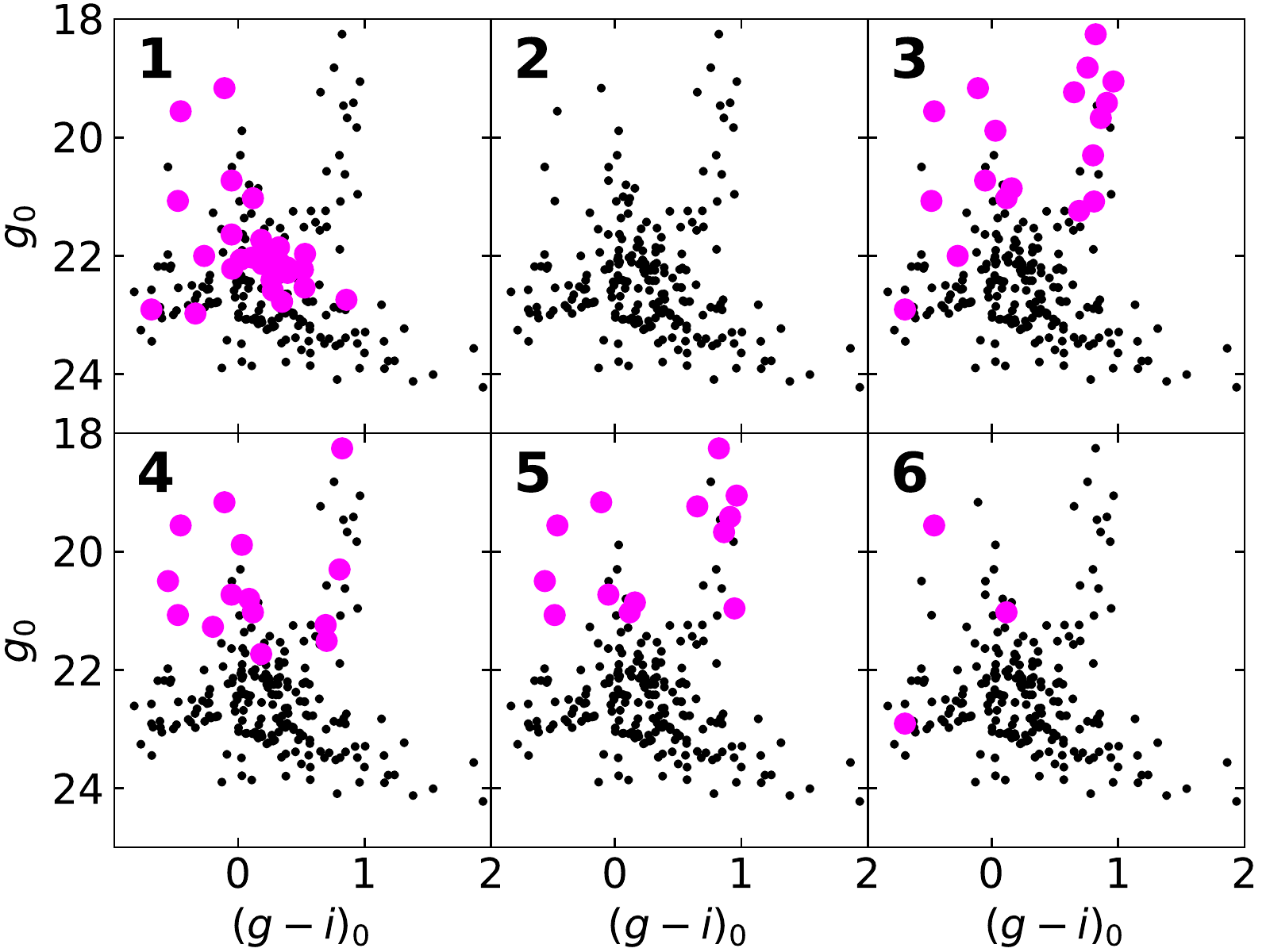}
\caption{Color-magnitude diagram of STEP-0029. Black points represent all the
measured stars in SMASH DR2 data sets located within a circle with a radius
equal to 3 times the cluster radius. Large magenta points represent the
stars that remained unsubtracted after the CMD cleaning procedure. The
reference star field used to decontaminate the star cluster CMD is indicated
at the top-left margin (see also Fig.~\ref{fig1}).}
\label{fig2}
\end{figure}

\begin{figure}
\includegraphics[width=\columnwidth]{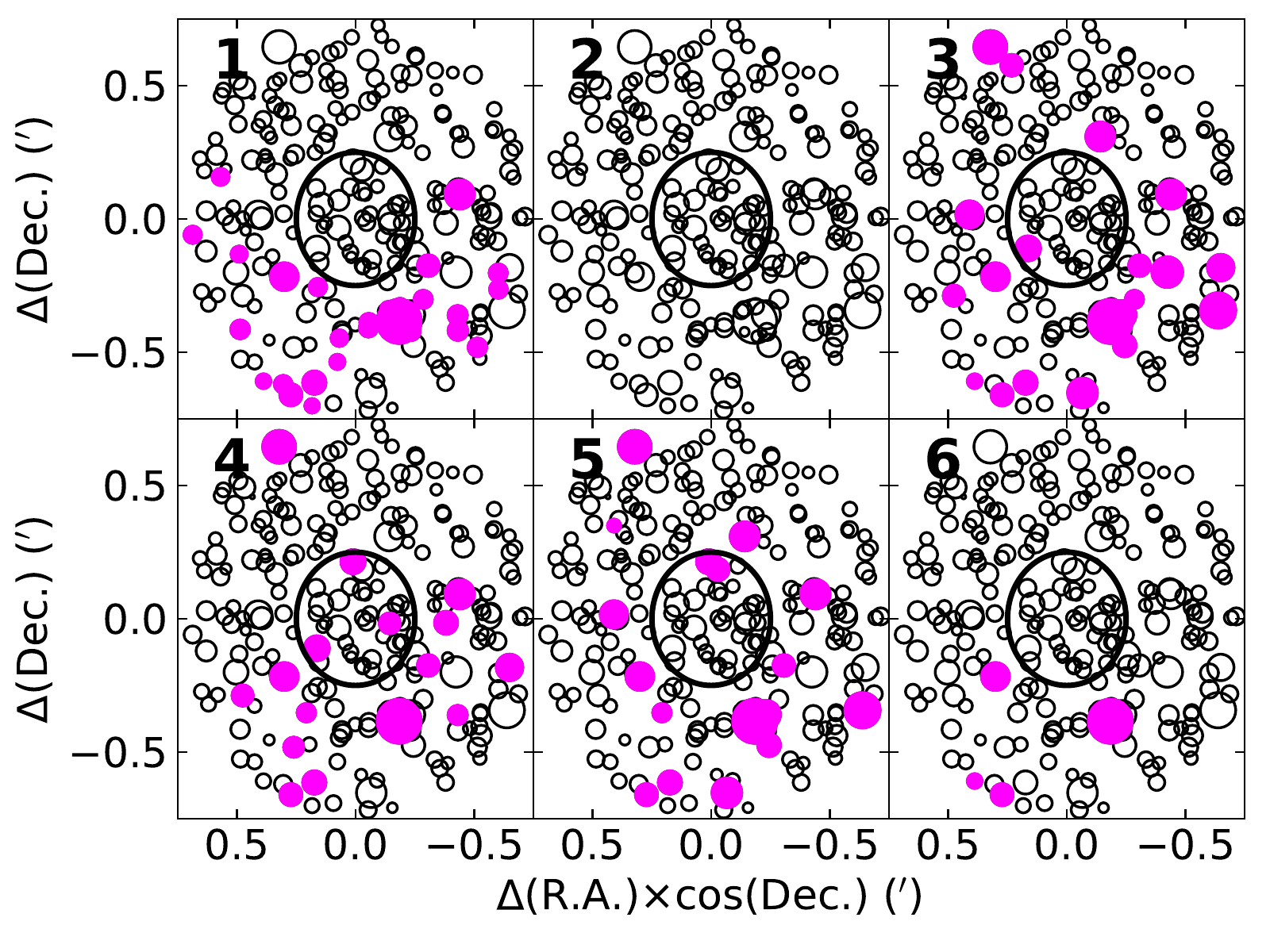}
\caption{Chart of the stars in the field of STEP-0029. The size of the symbols
is proportional to the $g$ brightness of the star. Open black circles
represent all the measured stars in SMASH DR2 data sets located within a circle 
with a radius equal to 3 times the cluster radius. Filled magenta circles represent the
stars that remained unsubtracted after the CMD cleaning procedure. The
reference star field used to decontaminate the star cluster CMD is indicated
at the top-left margin (see also Fig.~\ref{fig1}). The large centered circle
represents that of the star cluster radius.
}
\label{fig3}
\end{figure}

Figure~\ref{fig2} illustrates the  different results of the
decontamination of field stars  when the different 6  reference star fields  (see Fig.~\ref{fig1})
are used, {\bf separately.} As can be seen, the  different resulting cleaned star cluster CMDs (magenta points)
show distinct groups of stars, depending on the reference star field used, which 
suggests that differences in the astrophysical properties of the composite star field population do exist. 
 If all the reference star fields showed a uniform distribution of stars in magnitude and color, all
the resulting cleaned CMDs should look similar.
The spatial distribution of  the stars that remained unsubtracted is shown in Fig.~\ref{fig3}.
From Figs.~\ref{fig2} and \ref{fig3} is readily visible that the stars that have
survived the cleaning procedure are not spatially distributed inside the cluster
radius  (black circle), nor they unquestionably follow the  expected sequences in the star cluster
CMD either. This means that those stars could rather represent fluctuations in the
stellar density along the LOS of the composite stellar field population.

We finally  assigned a membership probability to each star that remained unsubtracted
after the decontamination of the star cluster CMD. Because the stars in the cleaned
CMDs vary with respect to the reference star field employed  (see the distribution of
magenta points in Figs.~\ref{fig2} and \ref{fig3}), we  defined the probability
$P$ ($\%$) = 100$\times$$N$/6, where $N$ represent the number of time a
star was not subtracted during the six different CMD cleaning executions. With that
information on hand, we built Fig.~\ref{fig4}, which shows the spatial distribution
and the CMD of all the measured stars located in the field of STEP-0029. Stars with
different $P$ values were {\bf plotted} with different colors. We applied the above
cleaning procedure to the remaining 16 star cluster candidates discovered by 
\citet{gattoetal2020} with ages $\ga$ 4 Gyr, for which SMASH DR2 photometry
is available. The resulting SMASH cleaned star cluster CMDs and spatial distribution of the
measured stars are shown in Figs.~\ref{figa1} to \ref{figa3} of the Appendix, while
those from STEP/YMCA data sets are depicted in Figs.~\ref{figa4} to \ref{figa6} of the
Appendix.

\section{Discussion and conclusions}

\begin{figure}
\includegraphics[width=\columnwidth]{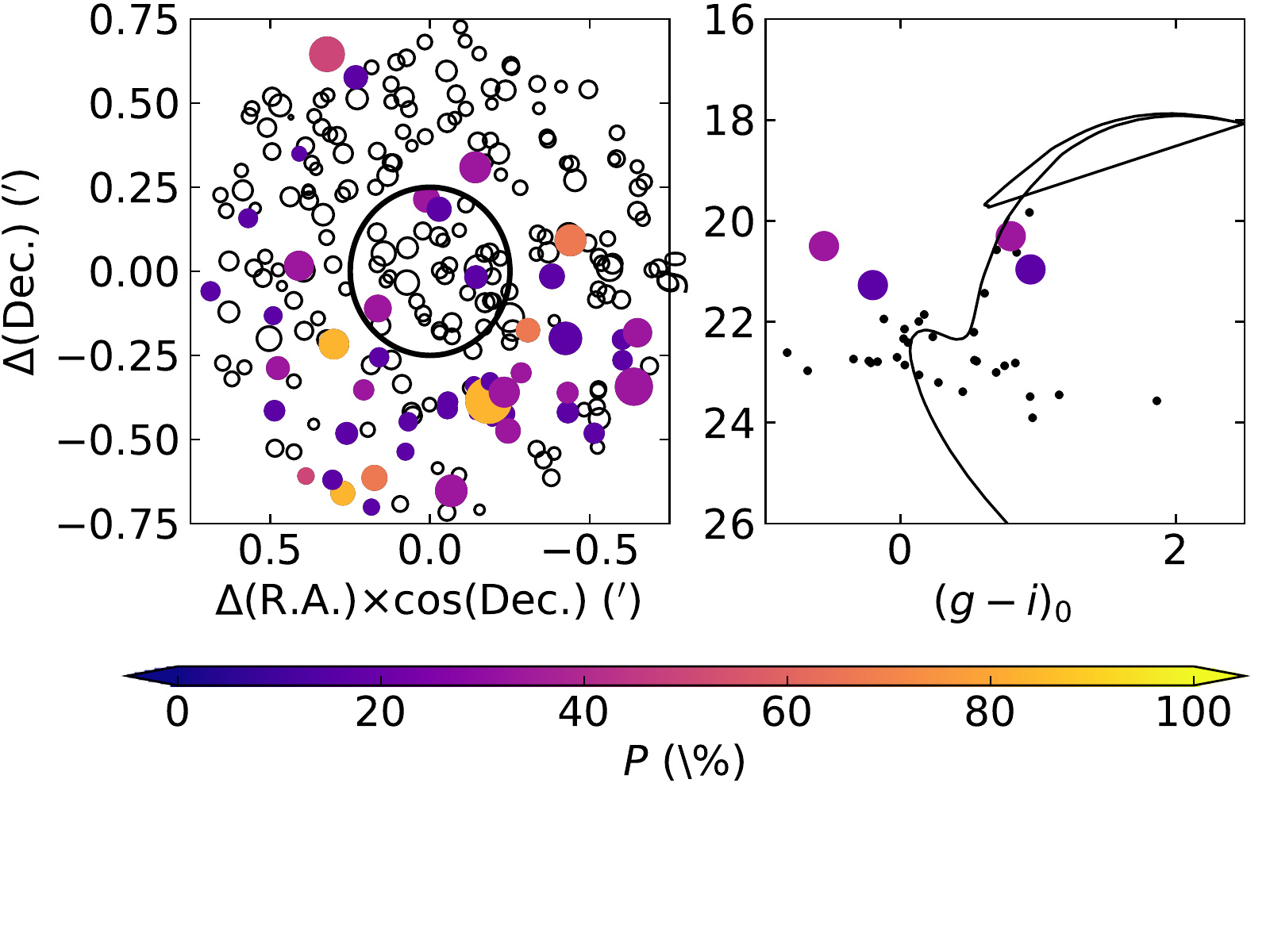}
\caption{{\it Left panel:}  Chart of the stars in the field of STEP-0029. The size of the symbols
is proportional to the $g$ brightness of the star. Open black circles
represent all the measured stars in SMASH DR2 data sets located within a circle 
with a radius equals to 3 times the cluster radius. The large
 centered circle represents that of the star cluster radius. {\it Right panel:}
Color-magnitude diagram of STEP-0029. Black points represent all the
measured stars in SMASH DR2 data sets located within the star cluster radius.
The theoretical isochrone plotted by \citet[][see their figure B1]{gattoetal2020}
is overplotted for comparison purposes.
Filled circles in both panels represent the
stars that remained unsubtracted after the CMD cleaning procedure, color-coded
according to the assigned membership probabilities ($P$).}
\label{fig4}
\end{figure}

By examining the spatial distributions of stars with assigned $P$ values
(color-coded symbols in Figs.~\ref{fig4}, \ref{figa1}-\ref{figa6}),
none of the analyzed fields show groups of stars with $P$ $>$ 50$\%$ 
concentrated inside the star cluster radius. This means that the
spatial overdensities discovered by \citet{gattoetal2020} do not highlight themselves 
in terms of stellar brightness and color distributions from those of the surrounding field
neither in the STEP/YMCA nor SMASH data sets, when the field star decontamination
 procedure described in Sect. 3 is applied. 
They could rather reveal small stellar density fluctuations in the studied LMC
regions.  Isolated stars spread throughout the cleaned areas spanning a wide range of 
$P$ values are seen in all the studied fields. We also detect some local concentrations 
of stars that distinguish from the surrounding field in the SMASH data, for example, toward the
southern outskirts of STEP-0012, YMCA-0021, and YMCA-0023. Their positions
in the cleaned CMDs, however, do not provide hints for any star cluster sequence.

Because of the LMC distance \citep[49.9 kpc;][]{dgetal14}, stars projected along a
particular LOS could produce CMDs with features similar to those seen in
star cluster CMDs. For example, from the SMASH cleaned CMD of YMCA-0002 
(Fig.~\ref{figa2}), we could conclude on the existence of a star 
cluster with some few red clump and main sequence turnoff stars, whereas the spatial 
distribution of stars with $P$ $>$ 70$\%$ does not support such a possibility. The 
SMASH cleaned CMD of YMCA-0007 shows a populous red clump of stars with $P$ $\sim$ 
50$\%$ that belong to the field, as judged by 
their spatial distribution. In summary, Figs.~\ref{fig4}, \ref{figa1}-\ref{figa6} most likely reveal the composite stellar population of the studied LMC regions and their local fluctuations.

\citet{p18d} arrived to a similar conclusion on the new identified star clusters
by \citet{bitsakisetal2017}, who found that the population of LMC star clusters
located at deprojected distances $<$ 4$\degr$ was nearly double the known
size of the system. \citet{p18d} based his findings on the remarkable large number of
objects with assigned ages older than 2.5 Gyr, which contrasts with the existence of 
the LMC star cluster age gap;  the fact that the assumption of a cluster formation 
rate similar to that of the LMC star field does not help to reconcile the large amount 
of star clusters either; and nearly 50$\%$ of them come from star cluster search 
methods known to produce more than 90$\%$ of false detections. 
\citet{bitsakisetal2017} identified only $\sim$ 35$\%$ of the previously known
cataloged LMC star clusters. The LMC star cluster frequency, i.e., number of star 
clusters per time unit, is a distribution function that basically does not 
change if low mass star clusters are not considered. The known LMC star cluster 
population is statistically complete down to 5$\times$10$^3$$M_\odot$ and
their star cluster frequency does not show clusters in the age-gap, and
this is a feature seen all throughout the LMC body \citep{p14b}. The LMC fields 
analyzed here are located beyond 4$\degr$ from the LMC center; the number of 
new detections is not such high; and the recovery fraction of known cataloged LMC
star clusters is much higher (see figure 7 in \citet{gattoetal2020}). Nevertheless, it is 
hardly possible that
age gap star clusters have been formed only in the outskirts of the LMC.
Indeed, the 15 ancient LMC globular clusters are distributed in the halo and in the disk of
the galaxy \citep{piattietal2019}.

The different outcomes obtained by \citet{gattoetal2020} and in the present work
reflect the different performances of the techniques employed for cleaning the CMDs of field
star contamination.  
When comparing the present constructed CMDs  (Figs.~\ref{fig4},
\ref{figa1}-\ref{figa6}) and those of \citet[][see their figure B1]{gattoetal2020} 
we note that: i) 
those built from SMASH and STEP/YMCA data contain  a similar number of stars and
reach in general  similar limiting magnitudes per unit area.  
 ii) By comparing the observed and cleaned CMDs built by
\citet{gattoetal2020}, it would seem that the number of field stars subtracted
was relatively small. Likewise, the number of stars considered as star cluster 
members  in \citet{gattoetal2020}'s CMDs would seem
also to be small as to conclude on clear features of a relatively old star cluster. 
In the present work, we subtracted a larger number of
field stars per unit area using the SMASH and STEP/YMCA data sets and no
definitive signature of star clusters are observed in the cleaned CMDs.
Therefore, we speculate with the possibility that \citet{gattoetal2020}'s
results  and ours are based on a low number statistics. 
iii) As can be seen, the star distribution along the theoretical 
isochrones in CMDs built from SMASH and  STEP/YMCA
data are comparable. Here we stress the issue that for most of the studied objects,
those stars would not seem to belong to a physical aggregate, 
but to the composite LMC field. Indeed,
the observed main sequence turnoffs are mostly populated by field stars. iv)
The metallicities of the isochrones used by \citet{gattoetal2020} 
(Z= 0.004, 0.006 and 0.008) are much metal-rich than the known metallicity of the
only one confirmed LMC age gap cluster ESO\,121-SC03  (Z $\approx$ 0.0015) \citep{pg13}.

Both independent studies point to the
need of further deeper observations of these star cluster candidates with the aim of
providing a definite assessment on their physical realities.
Our analysis shows that STEP/YMCA and SMASH surveys do not have 
the ability to disentangle the existence of LMC age gap star clusters.
The existence of LMC age gap star clusters has been
largely treated in the astronomical community, and the general consensus after different
focused campaigns for such star clusters and astrophysical results of the star formation 
history in the LMC has been that ESO 121-SC03 is the only known age gap star cluster.

\acknowledgements
 I thank the referee for the thorough reading of the manuscript and the
suggestions to improve it. 
I warmly thank Vincenzo Ripepi and his team who suggested changes in a previous version of this work.

This research uses services or data provided by the Astro Data Lab at NSF's National Optical-Infrared Astronomy Research Laboratory. NSF's OIR Lab is operated by the Association of Universities for Research in Astronomy (AURA), Inc. under a cooperative agreement with the National Science Foundation.



\begin{appendix} 

\section{Cleaned star cluster CMDs}

\begin{figure}
\includegraphics[width=6cm]{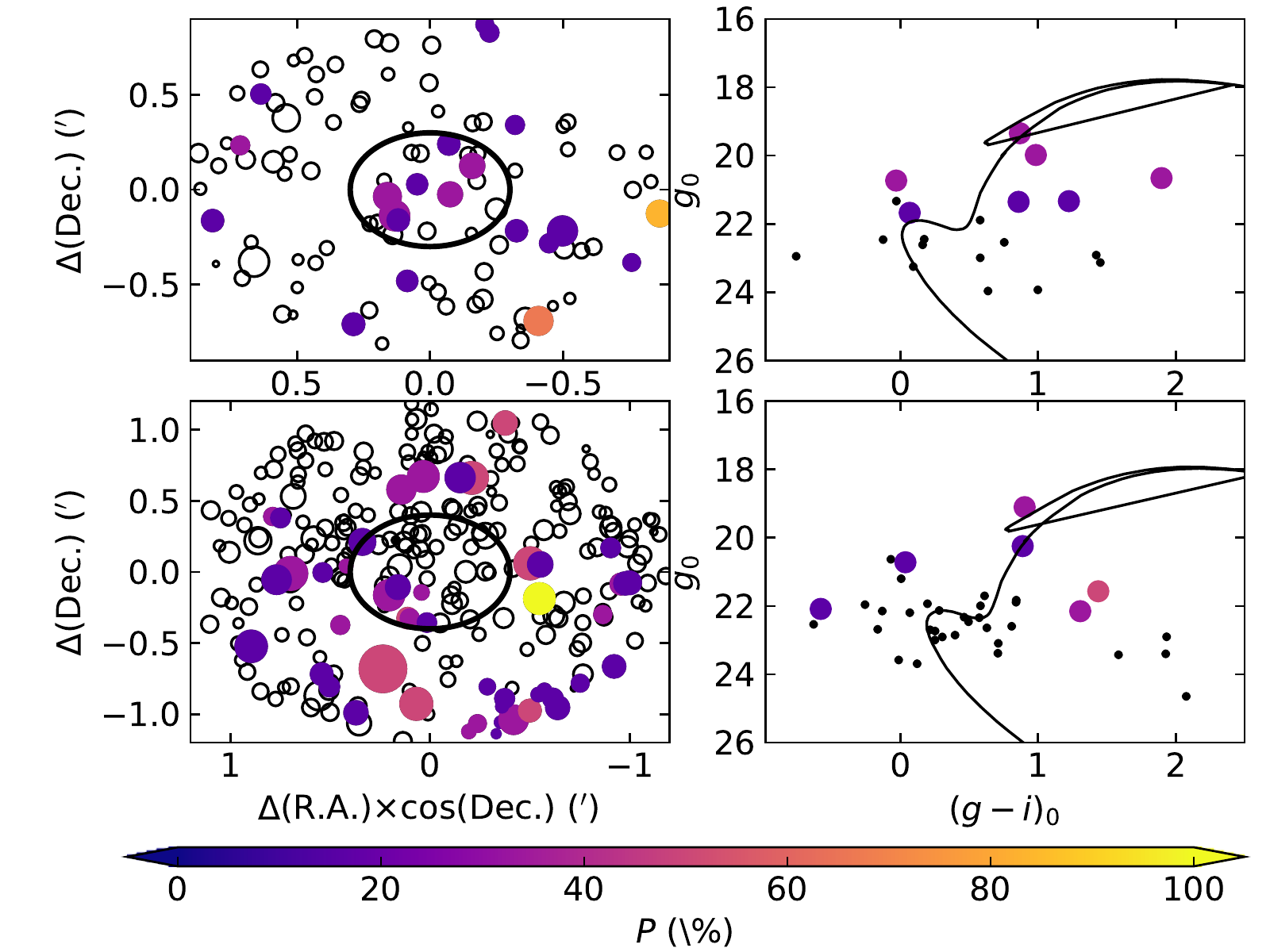}
\includegraphics[width=6cm]{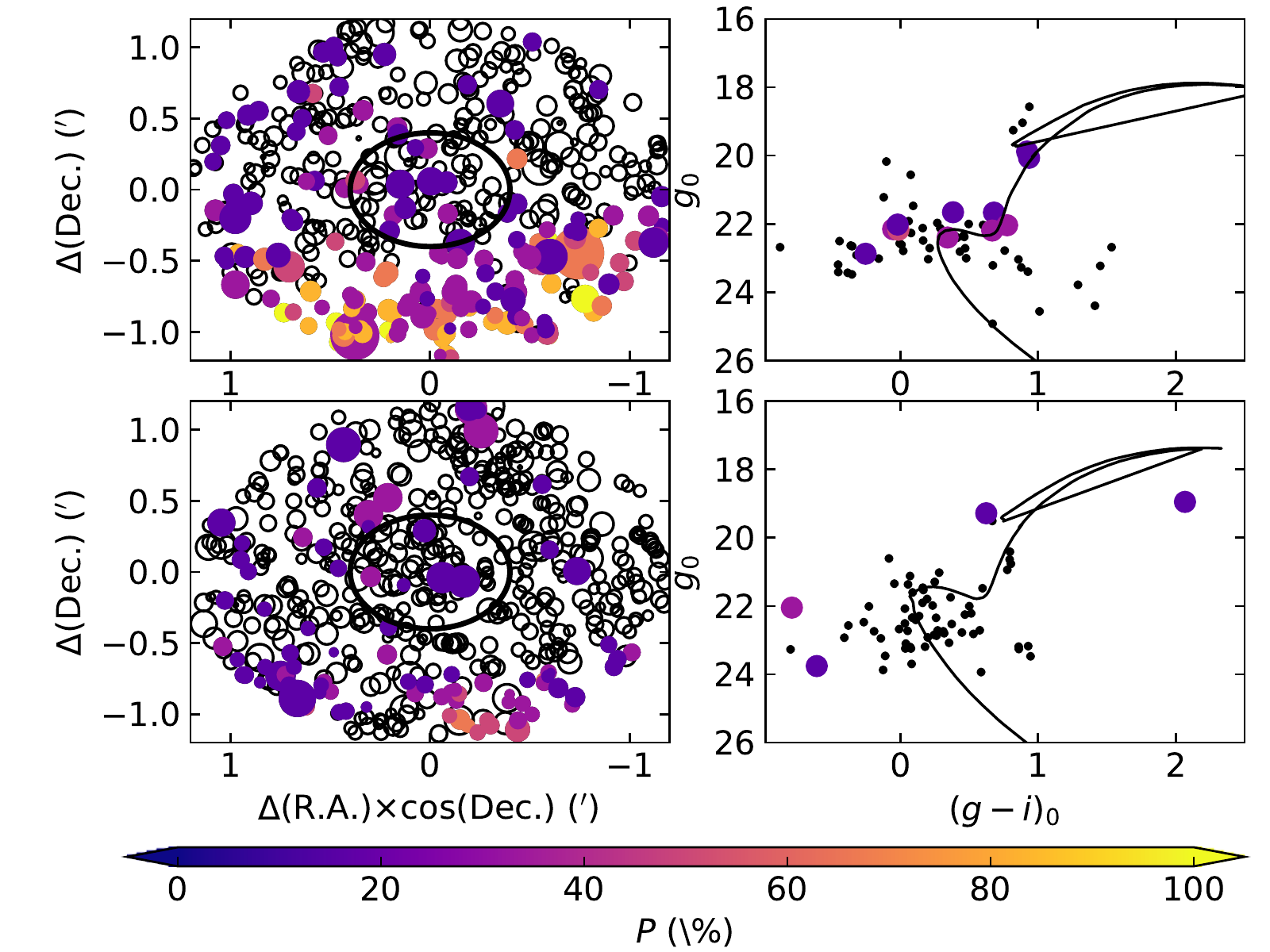}
\includegraphics[width=6cm]{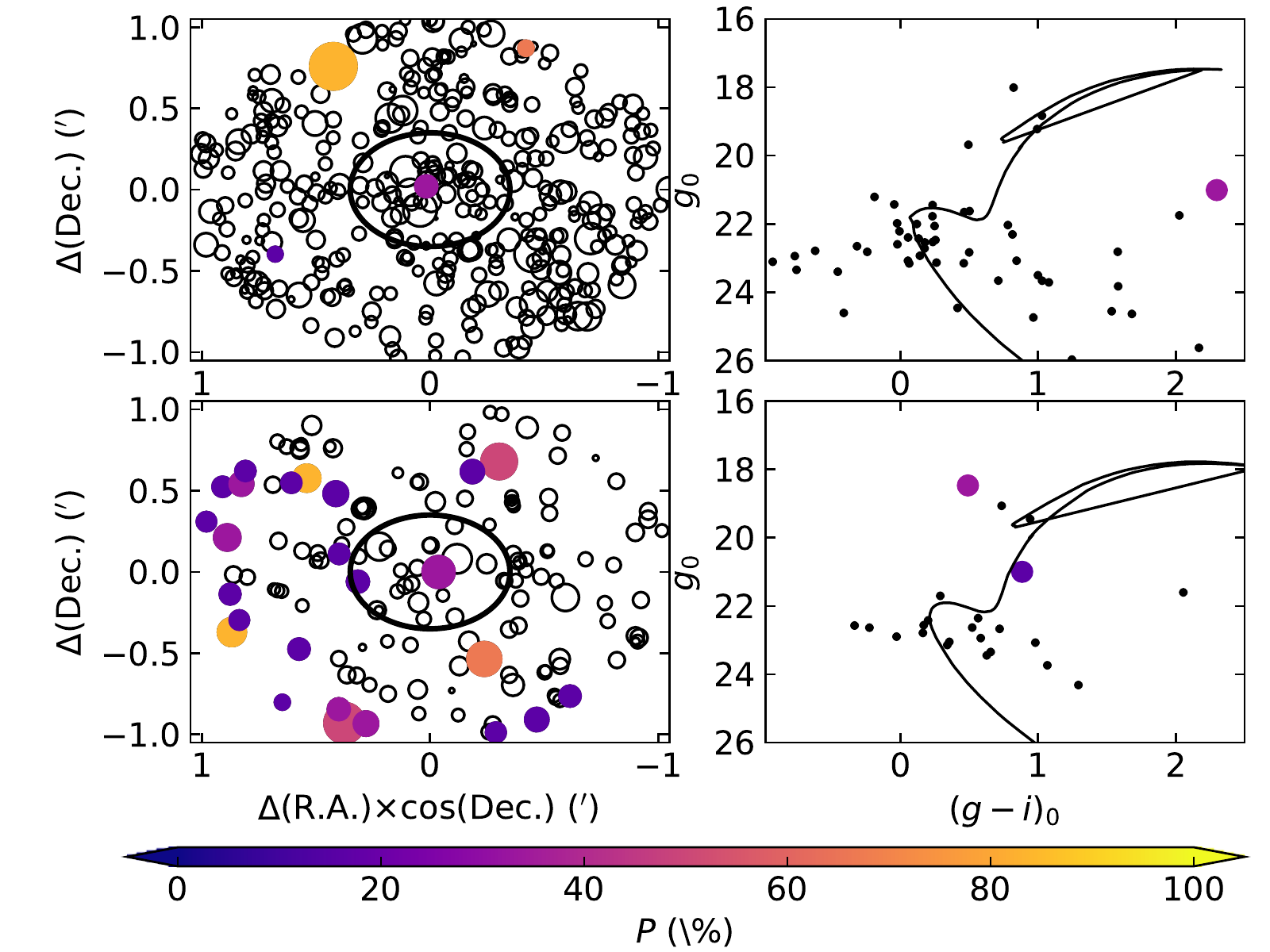}
\caption{Same as Fig.~\ref{fig4} for STEP-0001, STEP-0005, STEP-0012,
STEP-0024, STEP-0035, and YMCA-0001 from top bottom, and from left to right.}
\label{figa1}
\end{figure}

\begin{figure}
\includegraphics[width=6cm]{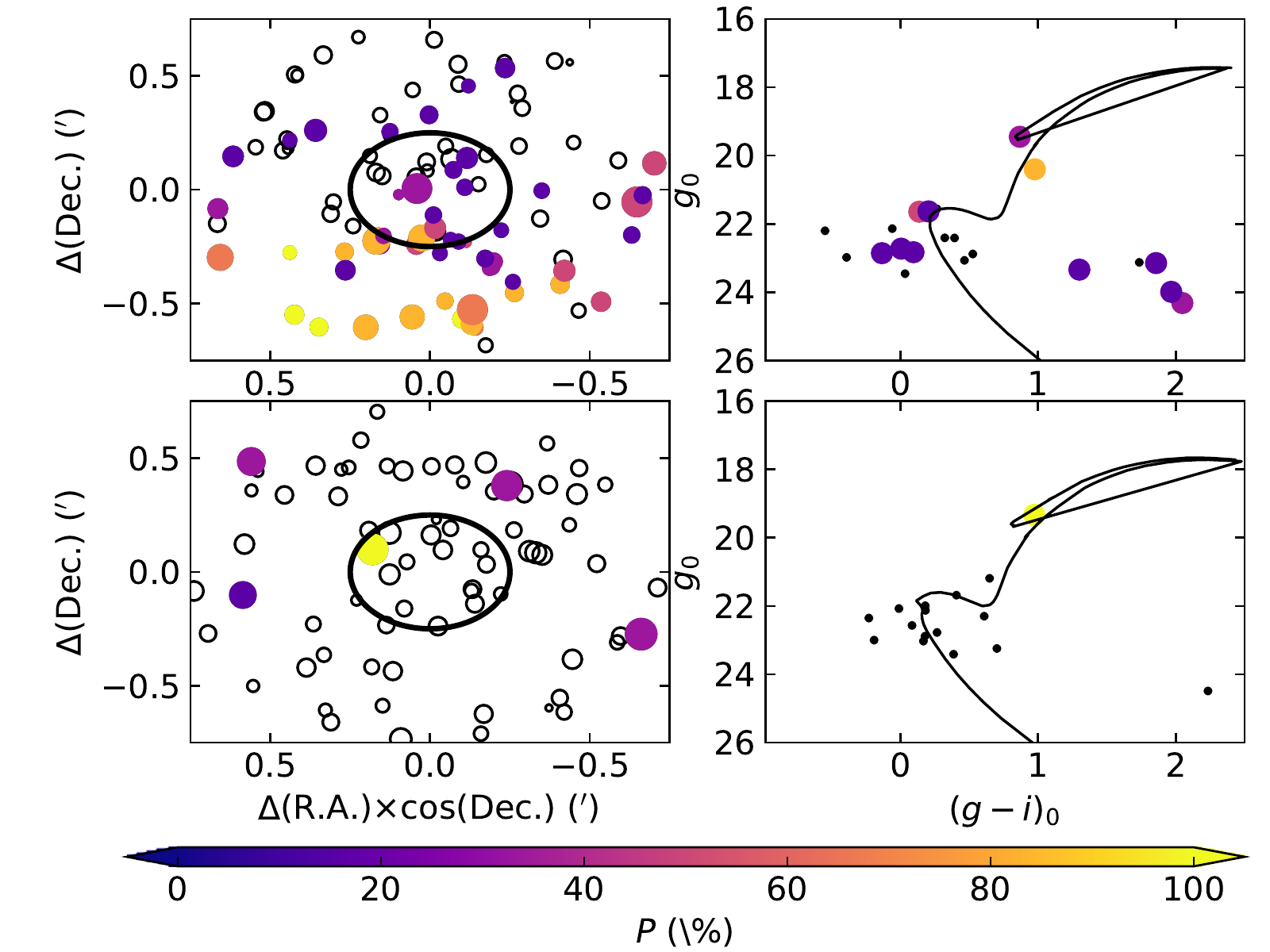}
\includegraphics[width=6cm]{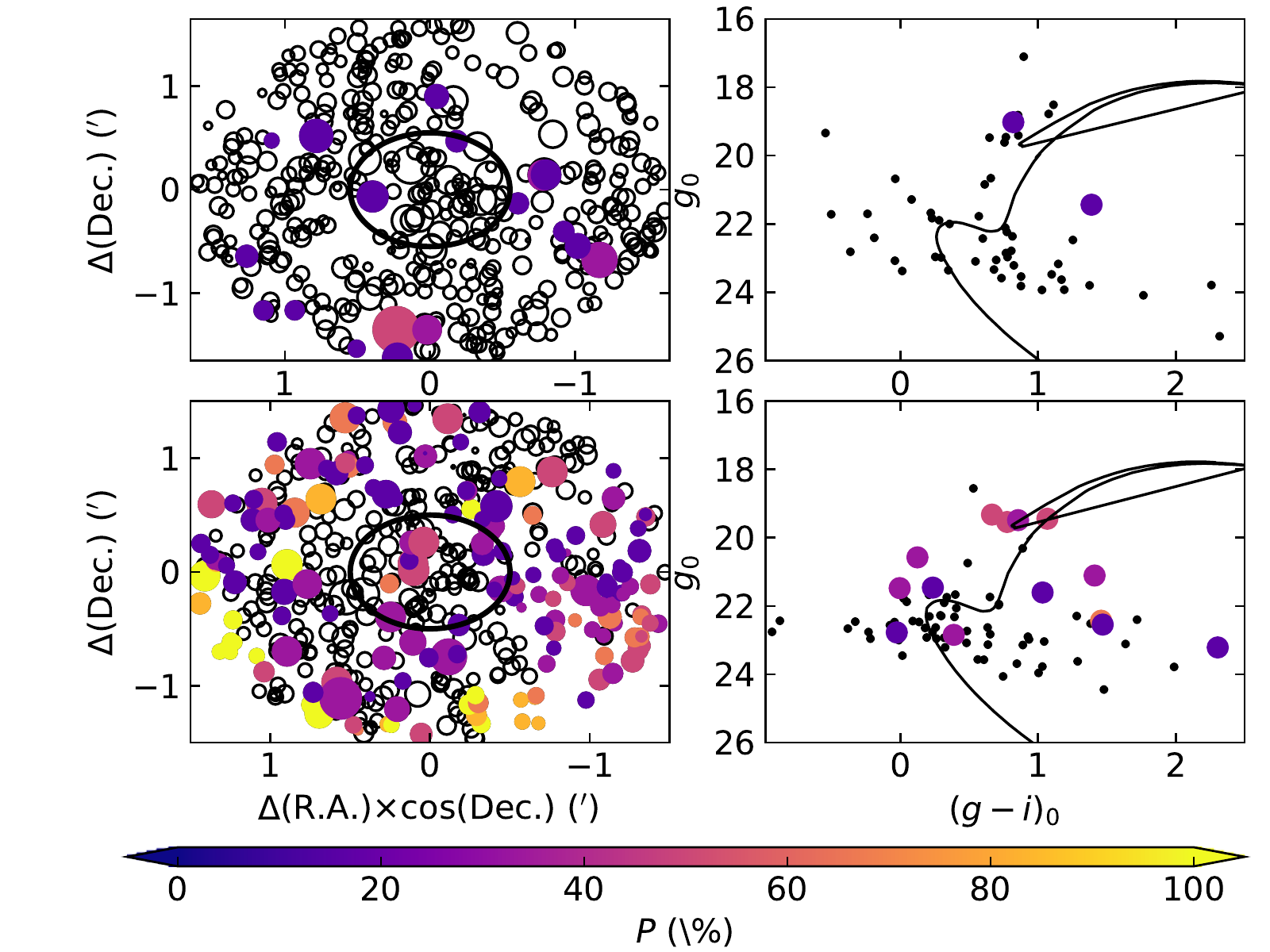}
\includegraphics[width=6cm]{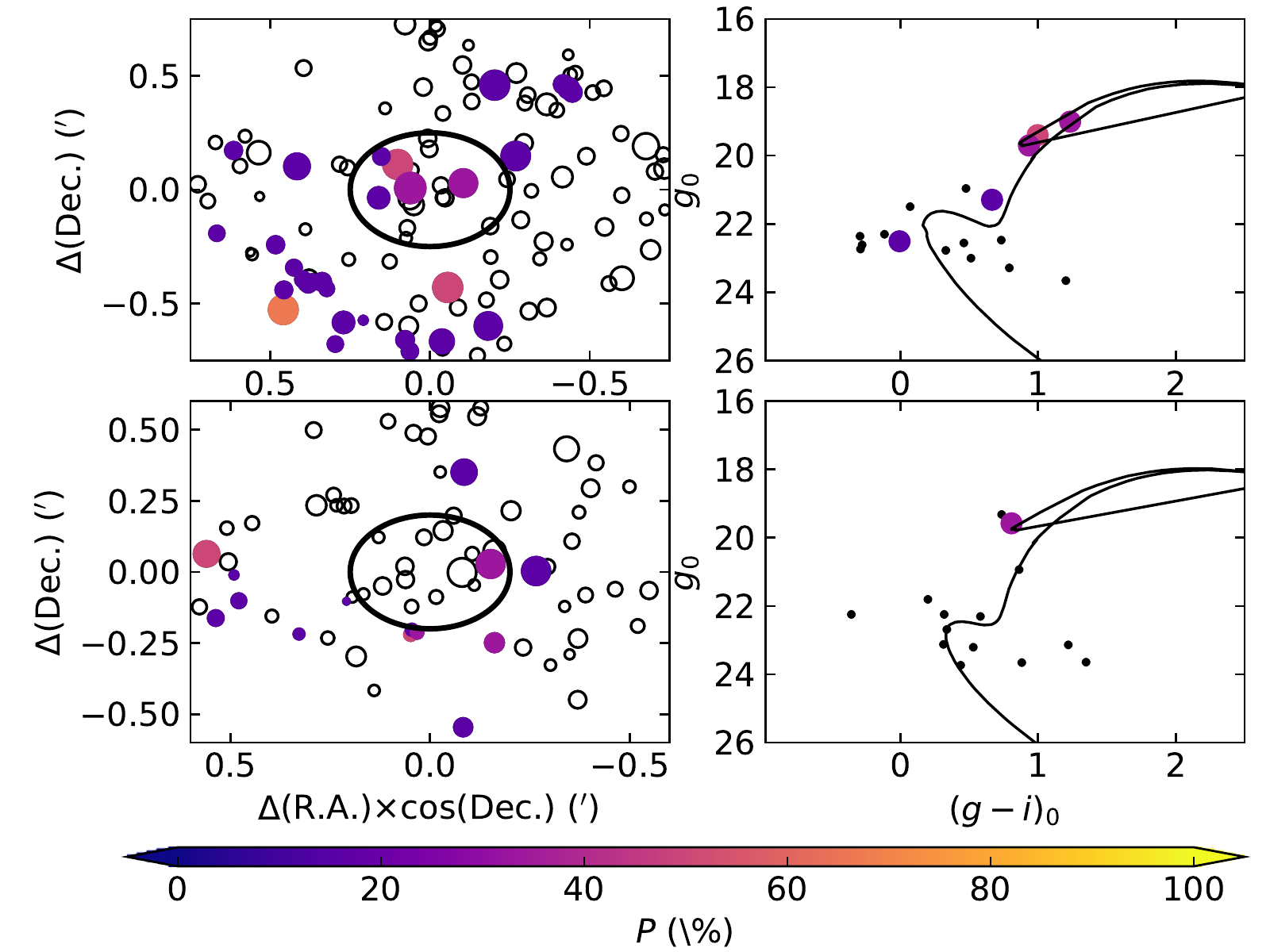}
\caption{Same as Fig.~\ref{fig4} for YMCA-0002, YMCA-0004, YMCA-0006,
YMCA-0007, YMCA-0008, and YMCA-0012 from top bottom,  and from left to right.}
\label{figa2}
\end{figure}

\begin{figure}
\includegraphics[width=6cm]{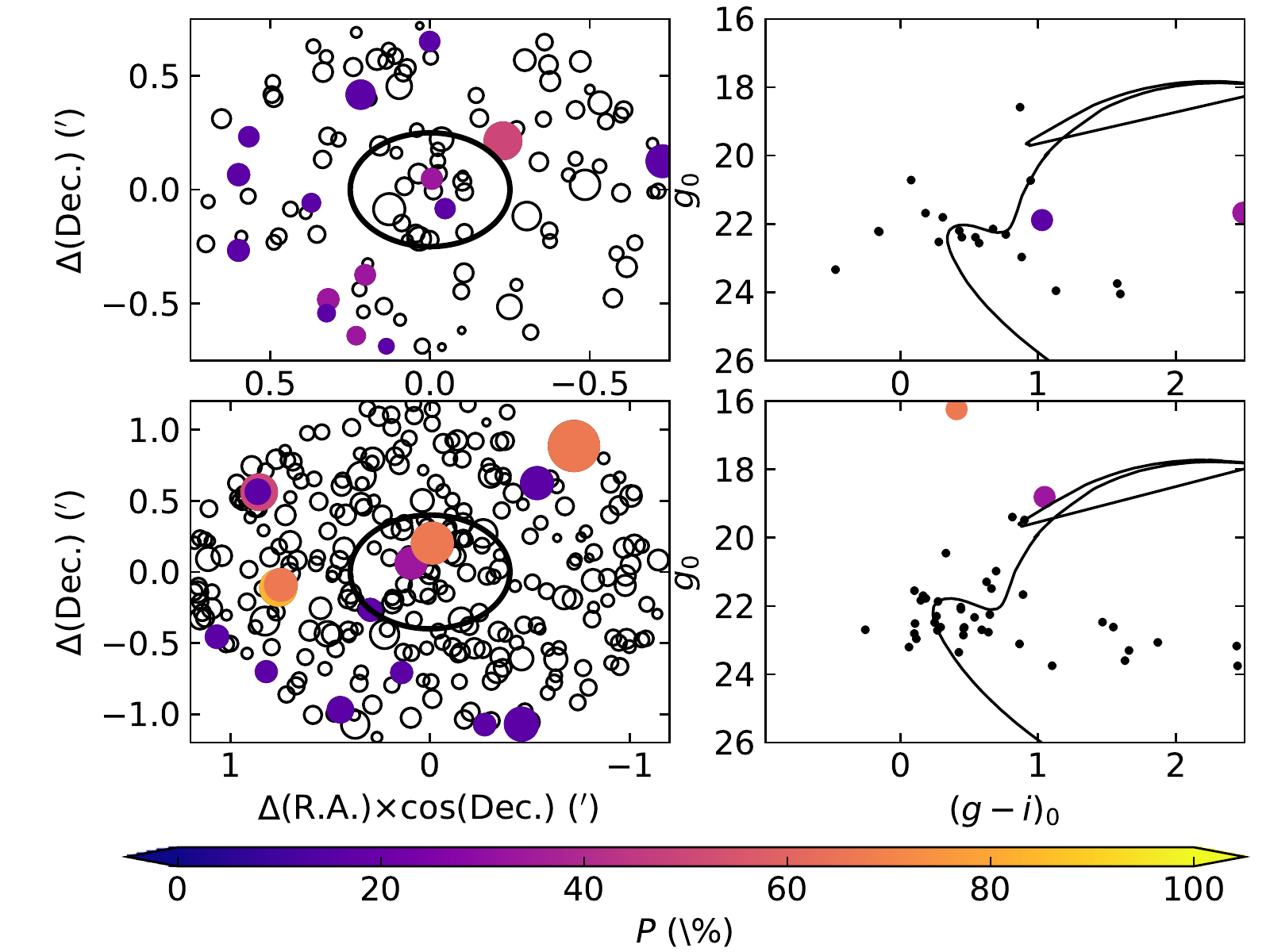}
\includegraphics[width=6cm]{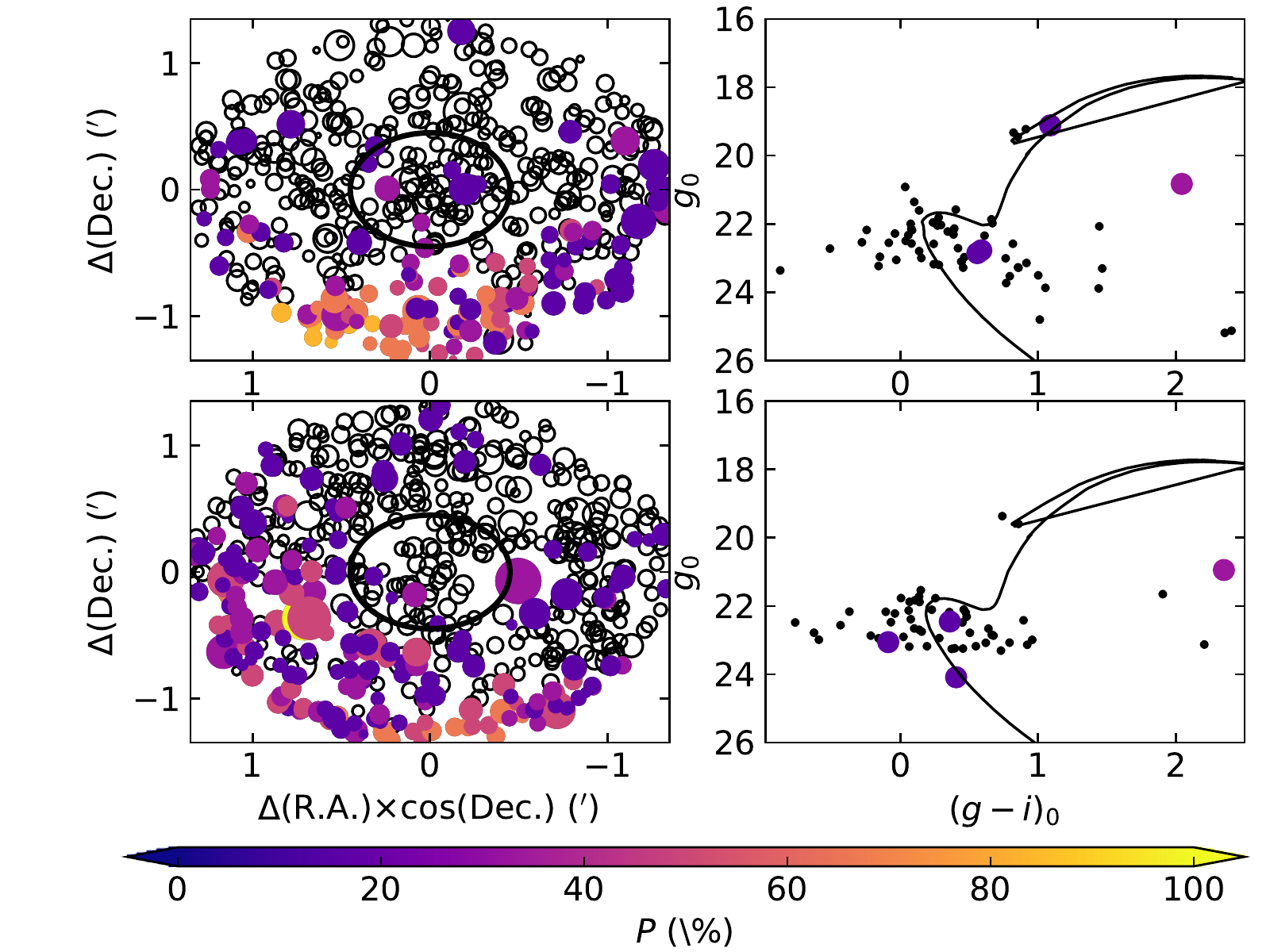}
\caption{Same as Fig.~\ref{fig4} for YMCA-0013, YMCA-0017, YMCA-0021,
and YMCA-0023 from top bottom,  and from left to right.}
\label{figa3}
\end{figure}

\begin{figure}
\includegraphics[width=6cm]{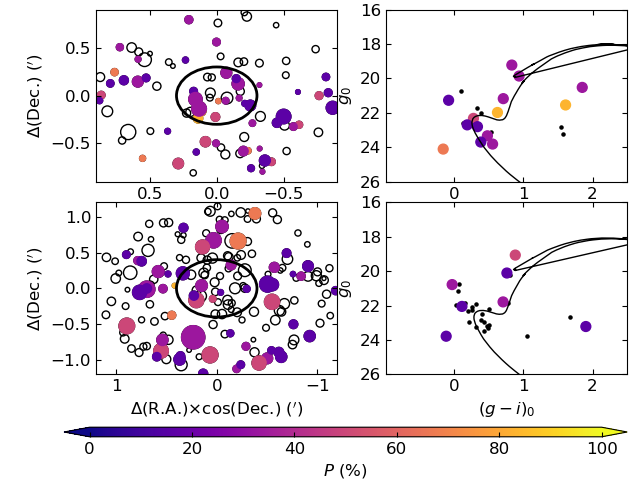}
\includegraphics[width=6cm]{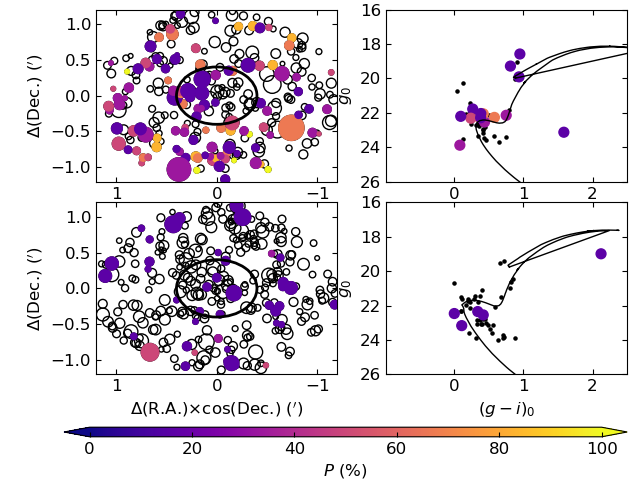}
\includegraphics[width=6cm]{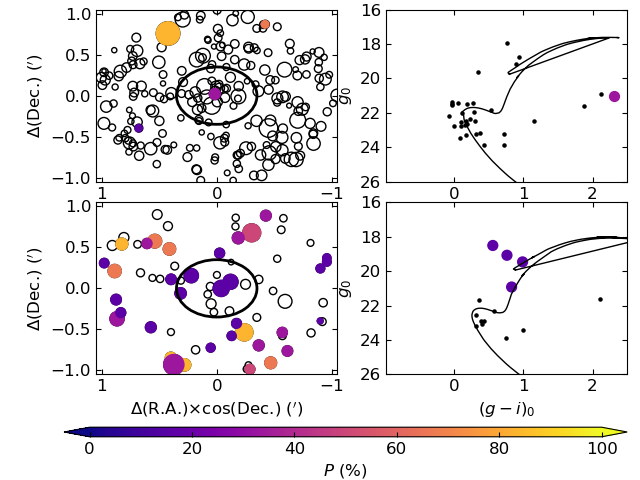}
\caption{Same as Fig.~\ref{fig4} for STEP-0001, STEP-0005, STEP-0012,
STEP-0024, STEP-0035, and YMCA-0001 from top bottom,  and from left to right. Data were kindly provided by V. Ripepi.}
\label{figa4}
\end{figure}

\begin{figure}
\includegraphics[width=6cm]{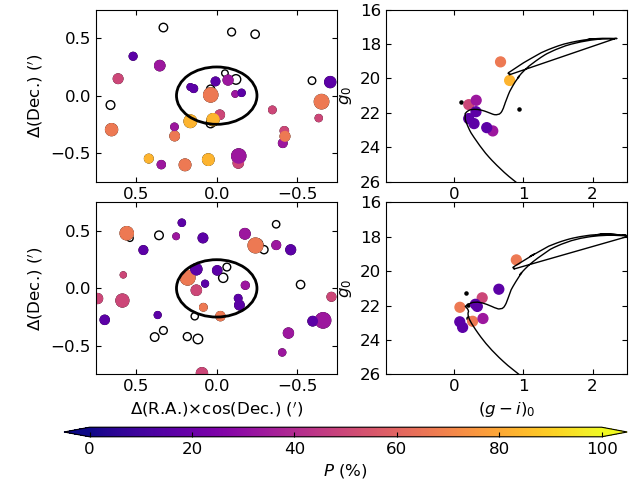}
\includegraphics[width=6cm]{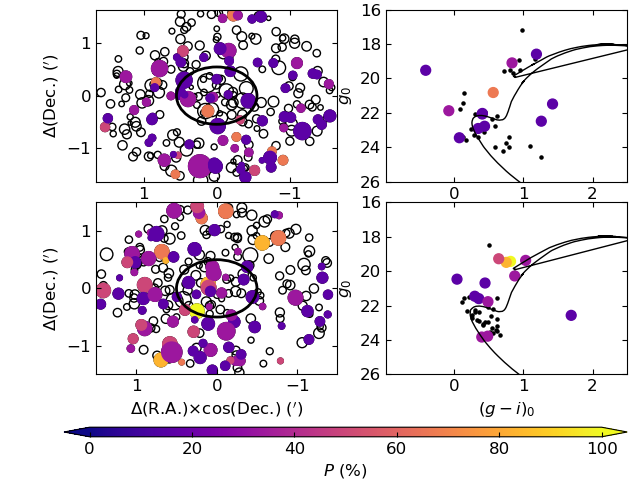}
\includegraphics[width=6cm]{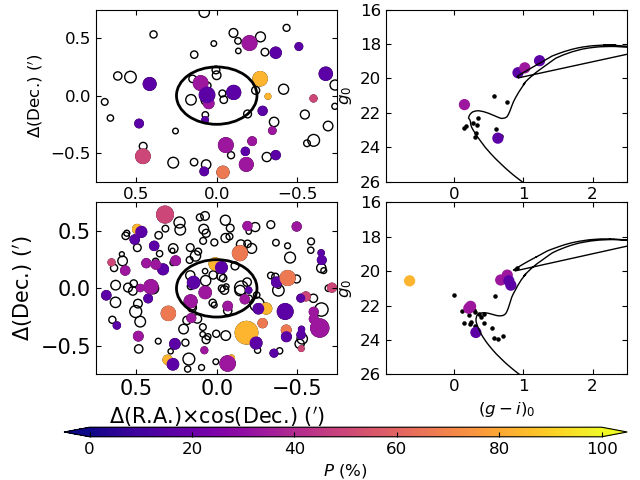}
\caption{Same as Fig.~\ref{fig4} for YMCA-0002,YMCA-0004,
YMCA-0006, YMCA-0007, YMCA-0008, and STEP-0029
 from top bottom,  and from left to right. Data were kindly provided
by V. Ripepi.}
\label{figa5}
\end{figure}

\begin{figure}
\includegraphics[width=6cm]{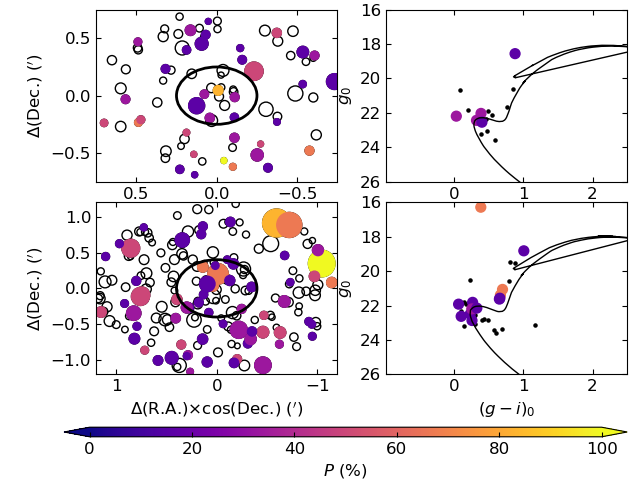}
\includegraphics[width=6cm]{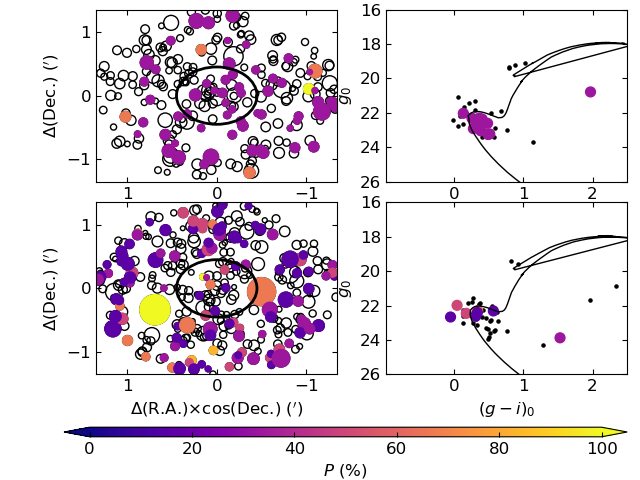}
\caption{Same as Fig.~\ref{fig4} for YMCA-0013, YMCA-0017, YMCA-0021,
and YMCA-0023 from top bottom,  and from left to right. Data were kindly provided
by V. Ripepi.}
\label{figa6}
\end{figure}

Figures~\ref{figa1} to \ref{figa3} show the SMASH cleaned star cluster
CMDs and the respective spatial distribution of the star candidates
with ages $\ga$ 4 Gyr discovered by \citet{gattoetal2020} . Symbols
are as in Fig~\ref{fig4}. Figures~\ref{figa4} to \ref{figa6} are those
produced from STEP/YMCA data sets.

\end{appendix}

\end{document}